# Electro-Thermal Transport in Metallic Single-Wall Carbon Nanotubes for Interconnect Applications

Eric Pop[1,2], David Mann[1], John Reifenberg[2], Kenneth Goodson[2] and Hongjie Dai[1]
[1]Laboratory for Advanced Materials, Chemistry and [2]Thermal Sciences Department, Mechanical Engineering
Stanford University, Stanford CA 94305, U.S.A.; Contact: epop@alum.mit.edu, hdai@stanford.edu

**Abstract**

This work represents the first electro-thermal study of metallic single-wall carbon nanotubes (SWNTs) for interconnect applications. Experimental data and careful modeling reveal that self-heating is of significance in short (1 < L < 10 μm) nanotubes under high-bias. The low-bias resistance of micron scale SWNTs is also found to be affected by optical phonon absorption (a scattering mechanism previously neglected) above 250 K. We also explore length-dependent electrical breakdown of SWNTs in ambient air. Significant self-heating in SWNT interconnects can be avoided if power densities per unit length are limited to less than 5 μW/μm.

**Introduction**

Metallic single-wall carbon nanotubes (SWNTs) have been proposed for interconnect applications owing to their ultra-high current density and insensitivity to electromigration [1,2]. Several recent studies have analyzed their potential as circuit elements [3,4,5]. However, no prior work has investigated their performance from a thermal point of view, an essential step in understanding their viability within integrated circuits. Despite their high thermal conductivity, the thermal *conductance* of carbon nanotubes is relatively low owing to their small diameter and thermal boundary resistance with the environment [6].

In this study we compare experimental data with a thermally self-consistent resistance model suitable for circuit analysis. For the first time, we find that the low-bias resistance of SWNTs with lengths relevant to interconnect applications (microns) increases with temperature due to optical phonon (OP) absorption above 250 K. We also find that shorter (1 < L < 10 μm) SWNT interconnects suffer comparatively more from self-heating than longer tubes for the same total power dissipated. Our results have significant implications for the viability of SWNT-based interconnects, and the proposed model covers a wide range of voltages and temperatures of practical relevance.

**Electrical Transport**

Figure 1 shows the SWNT layout considered in this work and Fig. 2 illustrates the *I-V* characteristics of a typical 3 μm long metallic tube. The solid lines in Fig. 2 are obtained with our temperature- and bias-dependent model, whereas the dashed lines represent the same model with the temperature fixed at the constant background value ($T_0$). The temperature dependence of the resistance is obtained through the temperature dependence of the electron scattering mean free paths (MFPs) with acoustic (AC) and optical (OP) phonons [7]. The total resistance is written as

$$R(V,T) = R_C + \frac{h}{4q^2}\left[\frac{L + \lambda_{eff}(V,T)}{\lambda_{eff}(V,T)}\right] \quad (1)$$

where $R_C$ is the electrical contact resistance and $\lambda_{eff}$ is the net, effective electron MFP

$$\lambda_{eff} = \left(\lambda_{AC}^{-1} + \lambda_{OP,ems}^{-1} + \lambda_{OP,abs}^{-1}\right)^{-1} \quad (2)$$

which includes electron scattering both by OP emission and absorption. We note that the latter has been neglected in previous studies due to the large OP energy in SWNTs (0.16–

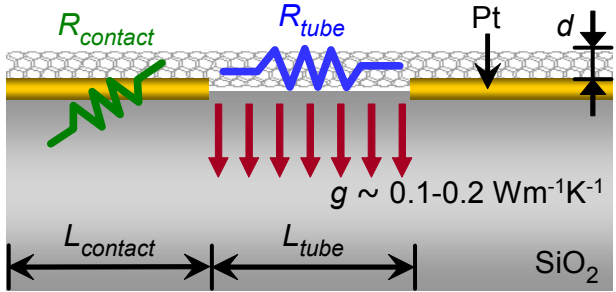

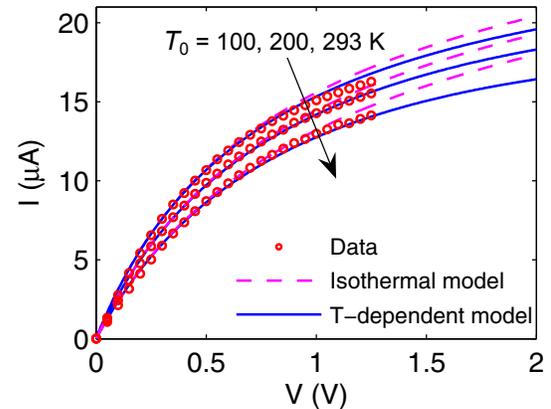

**Figure 1.** Schematic of two-terminal SWNT device used in the experiments and elements used for the analysis in this work. The thermal conductance from nanotube- and contacts-to-substrate is dominated by their interface, with $g \approx$ 0.1-0.2 WK$^{-1}$m$^{-1}$ per length.

**Figure 2.** Current-voltage (experimental data and model) of a metallic SWNT with $L \approx$ 3 μm and $d \approx$ 2.4 nm in ambient $T_0$ = 100, 200 and 293 K from top to bottom. The isothermal model deviates from the data beyond $V > 1$ V, suggesting SWNT self-heating.

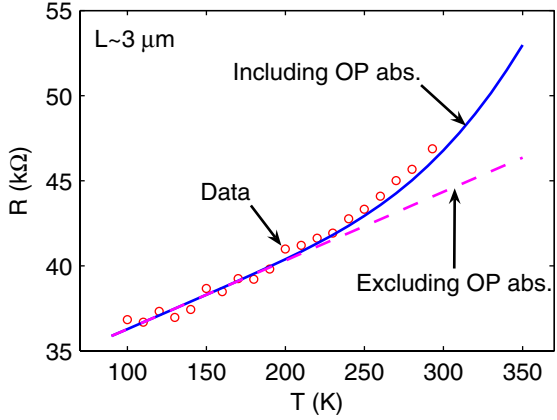

**Figure 3.** Temperature dependence of *low-bias* resistance for the SWNT from Fig. 2: data (symbols) and model with OP absorption (solid line) and without (dashed line). The importance of this scattering process is noted here for the first time, even at low bias for $T > 250$ K. The contact resistance was estimated $R_C \approx 24$ kΩ.

0.20 eV) and their relatively low occupation near room temperature [8,9,10]. This assumption is relaxed here, and scattering by OP phonon absorption is shown to play a non-negligible role at moderate to high temperatures, even at low applied bias (Fig. 3 and 5). The AC scattering and OP absorption lengths can be written respectively as [7,9]:

$$\lambda_{AC} = \lambda_{AC,300}(300/T) \quad (3)$$

$$\lambda_{OP,abs}(T) = \lambda_{OP,300}\frac{N_{OP}(300)+1}{N_{OP}(T)} \quad (4)$$

where $\lambda_{AC,300} \approx 1600$ nm is the AC scattering length at 300 K, and $\lambda_{OP,300} \approx 15$ nm is the spontaneous OP *emission* length at 300 K. We note that OP emission can occur both after electrons gain sufficient energy from the electric field, and after an OP absorption event:

$$\lambda_{OP,ems} = \left(1/\lambda_{OP,ems}^{fld} + 1/\lambda_{OP,ems}^{abs}\right)^{-1}. \quad (5)$$

The former MFP can be written as

$$\lambda_{OP,ems}^{fld}(T) = \frac{\hbar\omega_{OP}}{qV}L + \frac{N_{OP}(300)+1}{N_{OP}(T)+1}\lambda_{OP,300} \quad (6)$$

where the first term estimates the distance electrons must travel in the electric field ($F = V/L$) to reach the OP emission threshold energy ($\hbar\omega_{OP} \approx 0.18$ eV) [7,9], and the second term represents the temperature dependence of the spontaneous OP emission length. The OP emission MFP after an absorption event is obtained from Eq. 6 by replacing the first term with the OP absorption length of Eq. 4:

$$\lambda_{OP,ems}^{abs}(T) = \lambda_{OP,abs}(T) + \frac{N_{OP}(300)+1}{N_{OP}(T)+1}\lambda_{OP,300}. \quad (7)$$

This approach lets us express the temperature dependence of the relevant MFPs with respect to the acoustic and optical scattering lengths at 300 K. The simple method works because the scattering lengths scale as ratios of the phonon occupation terms for metallic SWNTs, whose density of states is nearly constant. The OP occupation is given by $N_{OP} = 1/[\exp(\hbar\omega_{OP}/k_BT) - 1]$. We note that in the limit of very small OP occupation (below room temperature) $N_{OP}(T)$ approaches zero, OP absorption can be neglected and the MFPs estimated above reduce to those of Refs. [8,9,10].

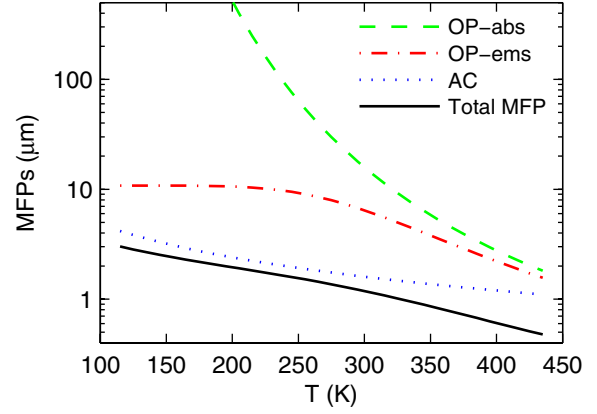

**Figure 4.** Estimate of the various MFPs at *low-bias* as a function of temperature. The total MFP ($\lambda_{eff}$ in Eq. 2) is also plotted (solid line). AC scattering dominates at low bias as expected, but OP scattering (with MFPs ≈ 10 µm at 300 K) cannot be neglected at higher temperatures, especially for longer tubes (see Fig. 5).

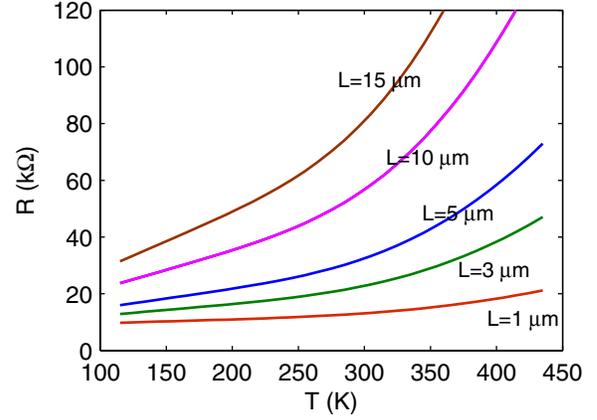

**Figure 5.** Computed temperature dependence of the *low-bias* resistance assuming ideal electrical contacts ($R_C = 0$) for various length tubes. The quantum contact resistance ($h/4q^2 \approx 6.5$ kΩ) is naturally still present (also see Figs. 3 and 4).

### Thermal Transport

The temperature profile along the SWNT depends on the power dissipated, and hence on its resistance. We obtain this temperature profile by solving the heat conduction equation along the length of the tube, including heat generation from Joule self-heating and heat loss to the substrate:

$$A\nabla(k_{th}\nabla T) + p' - g(T - T_0) = 0 \quad (8)$$

where $A = \pi db$ is the cross-sectional area ($b \approx 0.34$ nm the tube wall thickness), $k_{th}$ is the SWNT thermal conductivity, $p'$ $= I^2(R-R_C)/L$ is the Joule heating rate per unit length, and $g$ is the net heat loss rate to the substrate per unit length. In order to obtain the I-V curves, equations (1)-(8) are computed self-consistently along the length of the tube, and this iterative solution is repeated until the temperature converges within 0.1 K at each bias point. The current is then simply $I = V/R$, where $I$ and $R$ both depend on temperature. Here, we assume the OP population does not suffer severe non-equilibrium, unlike it was recently found for freely suspended tubes [7], since the presence of the substrate facilitates fast OP decay.

| | Thermal Resis. Phonons (K/W) | Thermal Resis. Electrons (K/W) | Electrical Resis. Electrons (kΩ) |
|---|---|---|---|
| Nanotube | 3 x 10$^8$ | 2 x 10$^9$ ← | ~ 15 |
| Contacts | 5 x 10$^6$ | 2 x 10$^9$ ← | ~ 15 |

**Figure 6.** Order of magnitude estimates for the resistance parameters of a metallic SWNT with $L \approx 2$ μm, $d \approx 2$ nm and contact $L_C \approx 2$ μm at room temperature. Arrows indicate estimates from the Wiedemann-Franz Law [15]. Phonons dominate heat conduction even along metallic nanotubes at room temperature and above [16].

The thermal conductivity of the SWNT is assumed to vary as $k_{th} \approx 3600(300/T)$ Wm$^{-1}$K$^{-1}$ above room temperature due to Umklapp phonon scattering [7]. However, since the dominant heat conduction pathway is downwards into the substrate, the results of this study are not too sensitive to thermal quantities that influence conduction along the tube (such as $k_{th}$ or $d$). We use a thermal conductance to the substrate $g \approx 0.15$ WK$^{-1}$m$^{-1}$ per tube length in our estimates for the small range of SWNT diameters examined here. This value is consistent with experimental estimates [11,12], with typical conductance values for solid-solid interfaces [13], and also with our own estimates from correlating the breakdown voltage and temperature of SWNTs (in Fig. 9 of this work). The thermal interface resistance between SWNT-substrate is estimated to be at least 10x greater than the thermal resistance of heat spreading into the substrate alone, so it dominates heat transfer away from the SWNT along its length.

## Comparison with Data and Predictions

The model is compared to data taken at several ambient temperatures ($T_0$) in Fig. 2. The dashed line in Fig. 2 illustrates the "ideal" scenario of perfect heat sinking by the substrate (isothermal conditions, SWNT temperature fixed at the background $T_0$), whereas the solid line takes Joule self-heating into account as described above. It is apparent that self-heating is already important for $V > 1$ V, although at much higher biases additional conduction channels and valleys may also come into play [10].

Figure 3 compares the *low-bias* (50 mV) SWNT electrical resistance from experimental data with our transport model including and excluding OP absorption over a wide temperature range. This demonstrates that, although previous neglected, OP absorption plays a non-negligible role at moderate to high temperatures ($T > 250$ K) and this scattering mechanism ought to be included in any future studies of SWNTs in this length range ($L > 1$ μm). Figure 4 displays the temperature dependence of all scattering mechanisms included in this study, from Eqs. 3-5. We estimate that the OP absorption length approaches 10 μm at room temperature, which explains its non-negligible role in longer nanotubes. The temperature dependence of low-bias resistance is further explored in Fig. 5 for several SWNT lengths, assuming perfect electrical and thermal contacts. Longer tubes have an earlier onset of OP phonon absorption, as is expected given the long OP absorption MFP (Fig. 4). The temperature coefficient of resistance (TCR) of metallic SWNTs around room temperature is $\approx 0.0026$ K$^{-1}$, which is very near that of 40 nm diameter copper vias [14].

Figure 6 summarizes the resistance components of a 2 μm single-wall nanotube with 2 μm long contacts. We use the Wiedemann-Franz Law [15] to estimate the contribution of electrons to thermal conduction, and find that it is negligible near room temperature and above (in accord with Ref. [16]). Finally, in order for the thermal resistance of the two contacts to play a minimal (less than 10 percent) role in the overall thermal resistance of the SWNT, we find that the length of the contacts must scale approximately as $L_C > 0.6/L$ (in μm). For a 2 μm SWNT this condition is satisfied if the contacts are at least 0.3 μm long, whereas for a 10 μm long SWNT the contacts are sufficiently long at 60 nm. This also serves to illustrate that for longer tubes heat is primarily dissipated directly into the substrate (also see Fig. 8), and contacts play a lesser role, as expected.

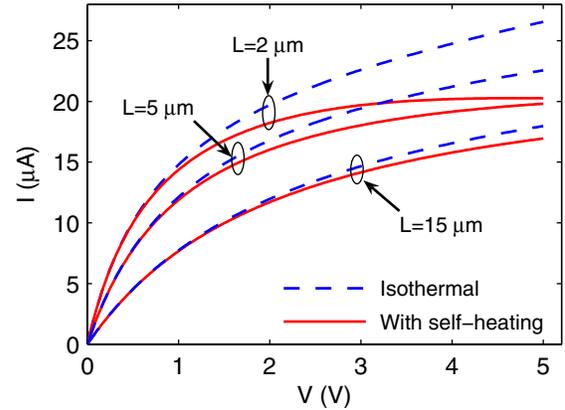

**Figure 7.** High-bias I-V characteristics computed with (solid lines) and without (dashed lines) taking self-heating into account. The ambient is $T_0 = 293$ K and the resistance parameters listed in Fig. 6 are used. Longer tubes show less self-heating (also see Fig. 8) due to lower power density and better heat sinking into the substrate.

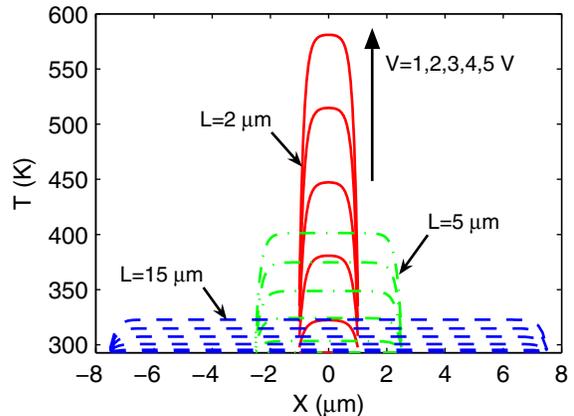

**Figure 8.** Estimated temperature profiles along the SWNTs from Fig. 7 when self-heating is included. The profiles are computed at $V = 1,2,3,4,5$ V from bottom to top (Eq. 8). The thermal healing length along the tube is $L_H \approx 0.2$ μm.

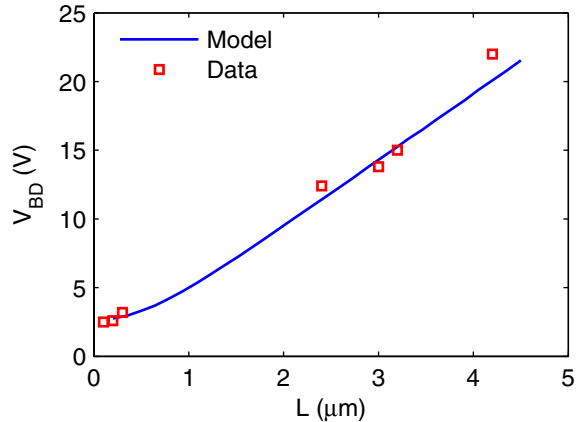

**Figure 9.** Computed and experimentally measured breakdown voltages ($V_{BD}$) for metallic SWNTs of various lengths exposed to air. The data was obtained in the course of this work and from Refs. [10,18]. Breakdown occurs by oxidation when the peak nanotube temperature (in their middle) reaches ≈ 800 K [17]. Inspection by AFM imaging revealed that tubes were indeed broken at or very near their midpoint, as expected [10].

We turn to self-heating at high bias in Figs. 7 and 8. Long ($L > 10$ μm) SWNTs do not show high temperature rise owing to lower power density and adequate heat sinking into the substrate along their length. However, we find that self-heating of SWNTs in the length range $1 < L < 10$ μm is significant under high bias and current conduction. This is due to the higher power density ($p' \approx IV/L$) and the large thermal resistances involved. In general, we find that power densities greater than 5 μW/μm will lead to noticeable self-heating of SWNTs in this length range, a simple rule which can be used as a design guideline. The assessment is more difficult for ultra-short ($L < 1$ μm) SWNTs, as more power is dissipated at the contacts, which also play a stronger role in heat-sinking. Future studies will address power dissipation issues in such ultra-short metallic nanotubes.

### High Bias Breakdown in Air

The temperature profile of an individual nanotube is very difficult to obtain experimentally in a quantitative manner [11]. However, the temperature under which SWNTs break down when exposed to air is relatively well established. Here, we exploit this mechanism to gain additional insight into the SWNT electrical and thermal properties, and the high temperature necessary for in-air breakdown is provided by self-heating under high bias. When exposed to air SWNTs break by oxidation which occurs at 800–850 K [17]. As it is sufficient for a single C-C bond to break in the single-wall tubes considered here, we choose the lower end of this range and assume breakdown occurs when the middle of the tube (point of highest temperature) reaches $T_{BD} \approx 800$ K. The scaling of the breakdown voltage ($V_{BD}$ in Fig. 9) is well reproduced by our model with 0.2 WK$^{-1}$m$^{-1}$ per tube length thermal conductance to the substrate. The breakdown voltage is found to scale approximately linearly with SWNT length (an experimentally well-documented finding [10,18]), also an indicator that longer tubes benefit from better heat sinking into the substrate along their length (the value of this thermal conductance being proportional to the length $L$). Although for interconnect applications SWNT bundles are most likely to be used [3], the analysis presented here remains the same and can be scaled accordingly.

### Conclusions

This work represents the first study of electro-thermal transport in metallic SWNTs relevant for practical interconnect applications ($L > 1$ μm). Electron scattering by optical phonon absorption was found to play a (previously neglected) role in altering the *low-bias* resistance of long SWNTs, while self-heating at high bias must be taken into account for shorter nanotubes. In addition, the breakdown voltage of SWNTs in air was found to scale linearly with the nanotube length. The resulting model has been validated against experimental data and is readily usable for circuit simulators or other design studies. Thermal management and design of high-current carrying nanotubes is expected to be important for future interconnect applications.

1. F. Kreupl *et al.* Proc. IEDM p. 683 (2004).
2. P. McEuen *et al.*, IEEE Trans. Nano. **1**, 78 (2002).
3. A. Naeemi *et al.*, IEEE EDL **26**, 84 (2005).
4. A. Raychowdhury and K. Roy, ICCAD p. 237 (2004).
5. N. Srivastava and K. Banerjee, VMIC p. 393 (2004).
6. C. Yu *et al.*, Nano Letters, **5**, 1842 (2005).
7. E. Pop *et al.*, Phys. Rev. Lett. **95**, 155505 (2005).
8. Z. Yao *et al.*, Phys. Rev. Lett. **84**, 2941 (2000).
9. J. Y. Park *et al.*, Nano Letters **4**, 517 (2004).
10. A. Javey *et al.*, Phys. Rev. Lett. **92**, 106804 (2004).
11. J. Small *et al.*, Solid State Comm. **127**, 181 (2003).
12. S. Huxtable *et al.*, Nature Materials **2**, 731 (2003).
13. D. Cahill *et al.*, J. Appl. Phys. **93**, 793 (2003).
14. W. Steinhogl *et al.*, Phys. Rev. B **66**, 075414 (2002).
15. C. Kittel, *Intro Solid-State Phys*, Wiley & Sons (1995).
16. T. Yamamoto *et al.*, Phys. Rev. Lett. **92**, 075502 (2004).
17. K. Hata *et al.*, Science **306**, 1362 (2004).
18. P. Qi *et al.*, J. Am. Chem. Soc. **126**, 11774 (2004).